\documentclass{dnce}
\usepackage{dnce}
\Page{1}           

\theoremstyle{definition}  
\theoremstyle{remark}      
\theoremstyle{plain}       

\theoremstyle{remark}

\theoremstyle{definition}

\titlemark{M. Belger, S. De Nigris, X Leoncini/ Discontinuity, Nonlinearity, and Complexity Vol-number (year) page1--page2}
\authormark{M. Belger, S. De Nigris, X Leoncini/ Discontinuity, Nonlinearity, and Complexity Vol-number (year) page1--page2}

\begin{document}

\title{\first{Slowing down of so-called chaotic states: ``Freezing'' the initial state }}

\setcounter{footnote}{1}

\author{\noindent\large ~M. Belger$^1$, S. De Nigris\footnote{\hspace*{2.5mm}email address:
denigris.sarah@gmail.com}~$^2$, X. Leoncini\footnote{Corresponding author, \hspace*{2.5mm}email address:
xavier.leoncini@cpt.univ-mrs.fr }~$^{1,3}$}


\address{\normalsize $^1$ Aix-Marseille Universit\'e, Universit\'e de Toulon, CNRS, CPT UMR 7332,
13288 Marseille, France \\ [-2mm]
$^2$ Department of Mathematics and Namur Center for Complex Systems-naXys,
University of Namur, \\ [-2mm]
8 rempart de la Vierge,
5000 Namur, Belgium  \\ [-2mm]
$^3$ Center for Nonlinear Theory and Applications, Shenyang Aerospace University,
Shenyang 110136, China }

\abstract{
\begin{table}[h!]
\vspace*{-5mm}
\doublerulesep 0.05pt
\tabcolsep 7.8mm
\vspace*{2mm}
\setlength{\tabcolsep}{7.5pt}
\hspace*{-2.5mm}\begin{tabular*}{16.5cm}{r|||||l}
\multicolumn{2}{l}{\rule[-6pt]{16.5cm}{.01pt}}\\
\parbox[t]{6cm}{\small
\vspace*{.5mm}
\hfill {\bf Submission Info}\par
\vspace*{2mm}
\hfill Communicated by Referees\par
\hfill Received DAY MON YEAR \par
\hfill Accepted DAY MON YEAR\par
\hfill Available online DAY MON YEAR\par
\noindent\rule[-2pt]{6.3cm}{.1pt}\par
\vspace*{2mm}
\hfill {\bf Keywords}\par
\vspace*{2mm}
\hfill Macroscopic Chaos\par
\hfill Hamiltonian Systems\par
\hfill Networks\par
\hfill Long-Range systems}
&
\parbox[t]{9.85cm}{
\vspace*{.5mm}
{\normalsize\bf Abstract}\par
\renewcommand{\baselinestretch}{.8}
\normalsize \vspace*{2mm} {\small The so-called chaotic states that emerge on the model of $XY$ interacting
on regular critical range networks are analyzed. Typical time scales
are extracted from the time series analysis of the global magnetization.
The large spectrum confirms the chaotic nature of the observable,
anyhow different peaks in the spectrum allows for typical characteristic
time-scales to emerge. We find that these time scales {$\tau(N)$}
display a critical slowing down, i.e they diverge as $N\rightarrow\infty$.
The scaling law is analyzed for different energy densities and the
behavior $\tau(N)\sim\sqrt{N}$ is exhibited. This behavior is furthermore
explained analytically using the formalism of thermodynamic-equations
of the motion and analyzing the eigenvalues of the adjacency matrix. }
\par
\par
\par
\hfill{\scriptsize\copyright 2012 L\&H Scientific Publishing, LLC. All rights reserved.}}\\[-4mm]
&\\
\multicolumn{2}{l}{\rule[15pt]{16.5cm}{.01pt}}\\\end{tabular*}
\vspace*{-7mm}
\end{table}
}

\maketitle
\thispagestyle{first}

\renewcommand{\baselinestretch}{1}
\normalsize

\section{Introduction}

Macroscopic chaotic behavior is often linked to out-of-equilibrium
states, one of the most prominent display of such phenomenon is most
certainly turbulence. The resulting chaotic or turbulent states result
from various macroscopic instabilities and bifurcations, and their
persistence is usually driven by strong gradients or energy fluxes.
When considering isolated systems with many degrees of freedom, some
similar behavior can be found, but typically it is a transient during
which, starting from a given initial condition, the system relaxes
to some thermodynamical equilibrium\cite{FPU55}. Microscopic ``molecular''
chaos plays there an important role for relaxation; however, in the
equilibrium state, macroscopic variables are at rest, despite the
microscopic chaos. It is nevertheless possible to extend this transient
state: indeed in recent years there has been an extensive study of
the so-called quasi-stationary states (QSSs), that emerge after a
violent relaxation in systems with long-range interactions \cite{Dauxois_book2002,Campa09,Campa_book2014,Levin2014}.
These states have the peculiarity that their lifetime diverges with
the number of constituents, so that the limits $N\rightarrow\infty$
and $t\rightarrow\infty$ do not commute. In fact it has been shown
that some of these states are non-stationary but can display regular
oscillations and, therefore, they represent a different kind of steady
state\cite{Holloway91,Vandenberg10,Yamaguchi2011,Ogawa2014}.
Moreover, as can be observed in \cite{Turchi2011}, both the lifetime of the state and
 the ``transient'' relaxation time from the QSS to the equilibrium
diverge with system size. During these relaxation periods we can expect
to observe some long lived but transient chaotic-like features in
these isolated states\cite{Antunes2015}. As mentioned,these transient
periods correspond to some kind of relaxation, nevertheless, more
recently, persistent chaotic macroscopic behavior in a isolated system
has been exhibited . These states occur over a wide range of energy.
They were first spotted on systems of rotators evolving on a regular
lattice, with a critical range of interaction and number of neighbors\cite{DeNigris2013a}.
Further studies have shown that this behavior occurred as well on
so-called lace networks, when the effective network dimension was
around $d=2$ \cite{DeNigris2015}. Studying these systems for different
number of constituents $N$ and a fixed density of energy $\varepsilon$,
the chaotic behavior of the order parameter was persistent, and it
appeared that the width of the fluctuations around its mean value
was not changing with $N$, implying an infinite susceptibility over
a given range of values of $\varepsilon$. However it was evident,
at least qualitatively, that some changes in the characteristic time
scales of the fluctuations were present and depended on the system's
sizes.

In this paper we focus on this dependence of the fluctuations time
scales with system size, we shall show that the observed scaling $\tau(N)\sim\sqrt{N}$
is different than the typical relaxation time scales observed in QSS,
and provide a theoretical explanation of these time scales in the
low energy range. The paper is organized as follows: in the first
part we describe the considered model and remind the reader of the
previously obtained results. We then move on to a a numerical study
of the characteristic time scales of the fluctuations by analyzing
the frequency spectrum of the measured order parameter, where a scaling
behavior $\tau(N)\sim\sqrt{N}$ is clearly exhibited. The presence
of a large and broad spectrum allows us to infer that the signal is
indeed chaotic. We then perform an analytical study of the thermodynamical
wave spectrum at low energies and we indeed confirm the numerically
exhibited scaling. This evidence confirms that these chaotic states
are not QSS's and that the chaotic behavior can be expected to be
an actual permanent feature or characteristic of these ``equilibrium''
states.

\section{Description of the model}

Originally the model we shall consider was tailored in order to uncover
the threshold of a long range interacting system. As such it was inspired
from the fact that the so-called $\alpha-$HMF model (see \cite{Anteneodo98, Campa01}) displayed
similar thermodynamical properties as the mean field model (for $\alpha<1$),
also dubbed the HMF model, which over the years has become de facto
the paradigmatic model to study and test new ideas when studying long
range system. In the $\alpha-$HMF rotators are located on a one-dimensional
lattice, and the coupling constant $J_{ij}$ between the spins decreases
according to a power-law with the distance between the rotators $J_{ij}\sim|i-j|^{-\alpha}$,
so that all rotators are coupled. The initial idea of the proposed
model was to consider a range $r$ of neighbors who equally interact
with a sharp edge, meaning that $J_{ij}\sim Cst$ if $|i-j|<r$ and
$0$ if $|i-j|\ge r$. We set up a window function, but what is important
here is that we allow $r$ to be a function of the total number of
spins $N$. The range is parametrized using a characteristic exponent
$1\le\gamma\le2$, which measures as well the total number of links
(interactions) being present in the system. When $\gamma=1$, we are
on a one-dimensional chain with a short range interactions (in our
case with just nearest  neighbors interactions), while when $\gamma=2$,
we retrieve the mean field model, with all rotators equally interacting
with each other. To get more specific we now present the details of
the rotators model placed on a one-dimensional lattice with periodic
boundary conditions. The Hamiltonian of the considered system writes 

\begin{equation}
H=\sum_{i=1}^{N}\frac{p_{i}^{2}}{2}+\frac{1}{2k}\sum_{i,j}^{N}\epsilon_{i,j}(1-\cos(q_{i}-q_{j}))\:,\label{eq: hamiltonian}
\end{equation}
where $k$ is the constant number of links (connections) per rotator
which scales with $\gamma$ as

\emph{
\begin{equation}
k\equiv\frac{1}{N}\sum_{i>j}\epsilon_{i,j}=\frac{2^{2-\gamma}(N-1)^{\gamma}}{N},\label{eq:degree}
\end{equation}
}and is related to the range by the simple relation $k=2r$. The matrix
$\epsilon_{i,j}$ is the adjacency matrix, defined as 

\begin{equation}
\epsilon_{i,j}=\begin{cases}
1 & if\,\,\left\Vert i-j\right\Vert \leqslant r\\
0 & otherwise
\end{cases}\:,\label{eq:adjacency matrix}
\end{equation}
where $\left\Vert i-j\right\Vert $ stands for the smallest distance
between two site on the one dimensional lattice with periodic boundary
conditions. From the Hamiltonian we directly get the equations of
the motion of the rotators.
\begin{eqnarray}
\dot{q}_{i} & = & p_{i}\label{eq:dynamics-1-1}\\
\dot{p}_{i} & = & -\frac{1}{k}\sum_{,j=1}^{N}\epsilon_{i,j}\sin(q_{i}-q_{j})\:.
\end{eqnarray}
A full study of the equilibrium properties of this model has been
made in \cite{DeNigris2013a,DeNigris2013b}. The order parameter that
we monitored is the total magnetization of the system $M$, defined
as 
\[
\mathbf{M}=\begin{cases}
M_{x}= & \frac{1}{N}\sum_{i=1}^{N}\cos q_{i}\\
M_{y}= & \frac{1}{N}\sum_{i=1}^{N}\sin q_{i}
\end{cases}=M\begin{cases}
\cos\varphi\\
\sin\varphi
\end{cases}
\]
The results are as follows:
\begin{figure}
\begin{centering}
\includegraphics[width=9cm]{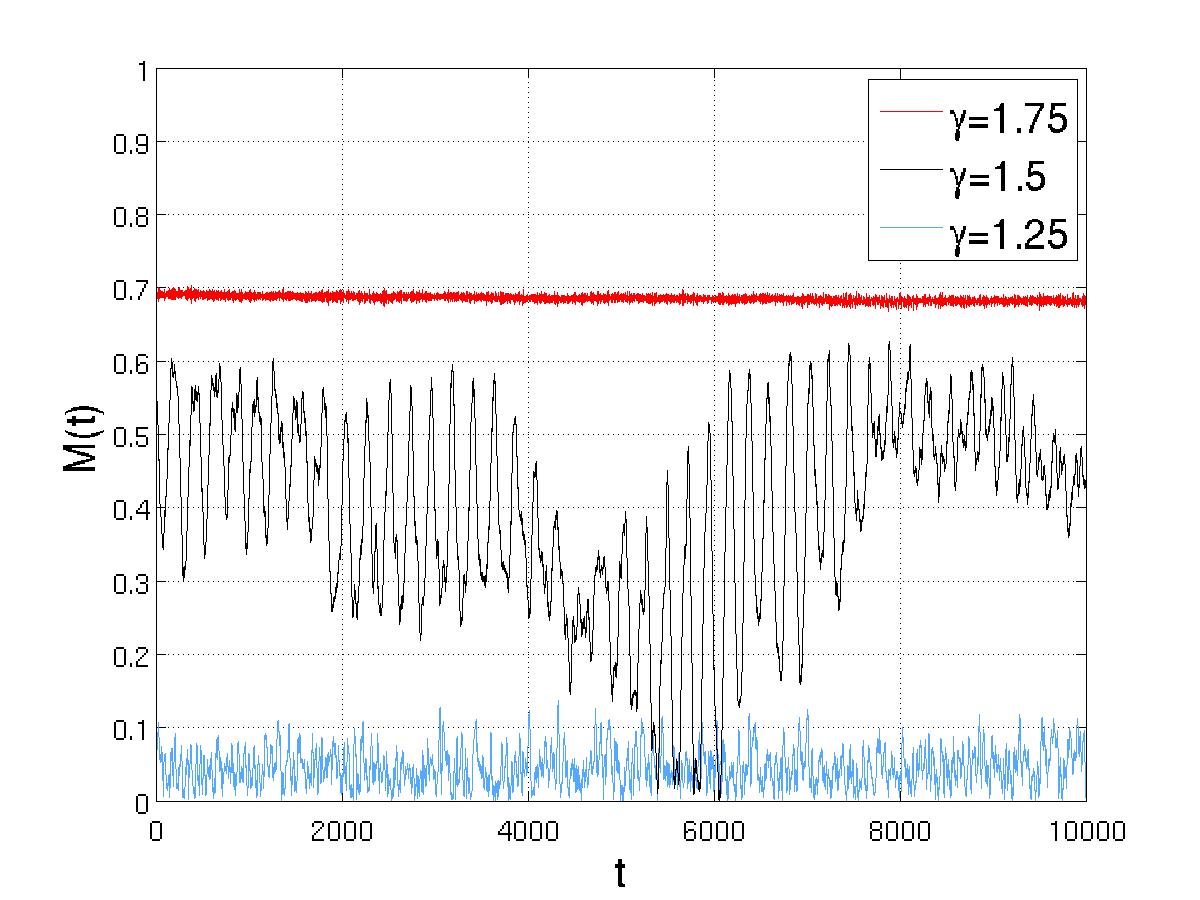}
\par\end{centering}

\centering{}\caption{Magnetization versus time, for a fixed density of energy $\varepsilon\approx0.4$
and different values of $\gamma$. The size of the system is $N=2^{13}$.
For $\gamma=1.25$ there is no magnetization (the residual magnetization
is due to finite size effects see for instance ), for $\gamma=1.75$
we observed a finite almost constant magnetization, while for $\gamma=1.5$
large fluctuations of order one are observed. Simulations have been
performed using a time step $\delta t=10^{-3}$. \label{fig:Fluctuations_Gamma} }
 
\end{figure}

\begin{itemize}
\item For $\gamma<1.5$ the system behaves as a short range model, meaning
that no order parameter emerges in the thermodynamic limit and no
phase transition exists. For the short range case ($\gamma=1$), this
result is consistent with the predictions of the Mermin-Wagner theorem,
which predicts that no order parameter can exist for systems with
dimensions $d\le2$, due to the existence of a continuous symmetry
group (here the global translation/rotation symmetry $q_{i}\rightarrow q_{i}+\theta$).
\item For $\gamma>1.5$ the system behaves like the mean field model, meaning
that a second order transition at a critical density of energy of
$\varepsilon_{c}=0.75$, is observed. All curves $M_{\gamma}(\varepsilon)$
appear as independent of $\gamma$ and fall on the mean field one.
\item For $\gamma=1.5$ for a range of temperatures below the critical energy
one, a chaotic state is observed. The magnetization displays steady
and large incoherent fluctuations, which do not appear to be dependent
on system size, implying an infinite susceptibility. The time dependence
of these fluctuations is the subject of this paper. Note also that
the transition of the Berezinsky-Kosterlitz-Thouless type has not
been detected (see for details \cite{DeNigris2013a}).
\end{itemize}
To illustrate the phenomena described, we have plotted in Fig.~\ref{fig:Fluctuations_Gamma}
the evolution of the order parameter at a fixed density of energy
$\varepsilon$ for three different values of $\gamma$ and a system
size of $N=2^{13}$. Simulations have been performed using the optimal
fifth order symplectic integrator described in \cite{McLachlan92}, and the fast-Fourier
transform made use of the FFTW package. We can notice the peculiar
regime that appears for $\gamma=1.5$ where the magnetization displays
what looks like a macroscopic chaotic behavior.

In the next section we shall study in more detail the temporal behavior
of the order parameter in these chaotic states.

\section{Critical slowing down}

\subsection{Numerical study}

\begin{figure}
\begin{centering}
\includegraphics[width=9cm]{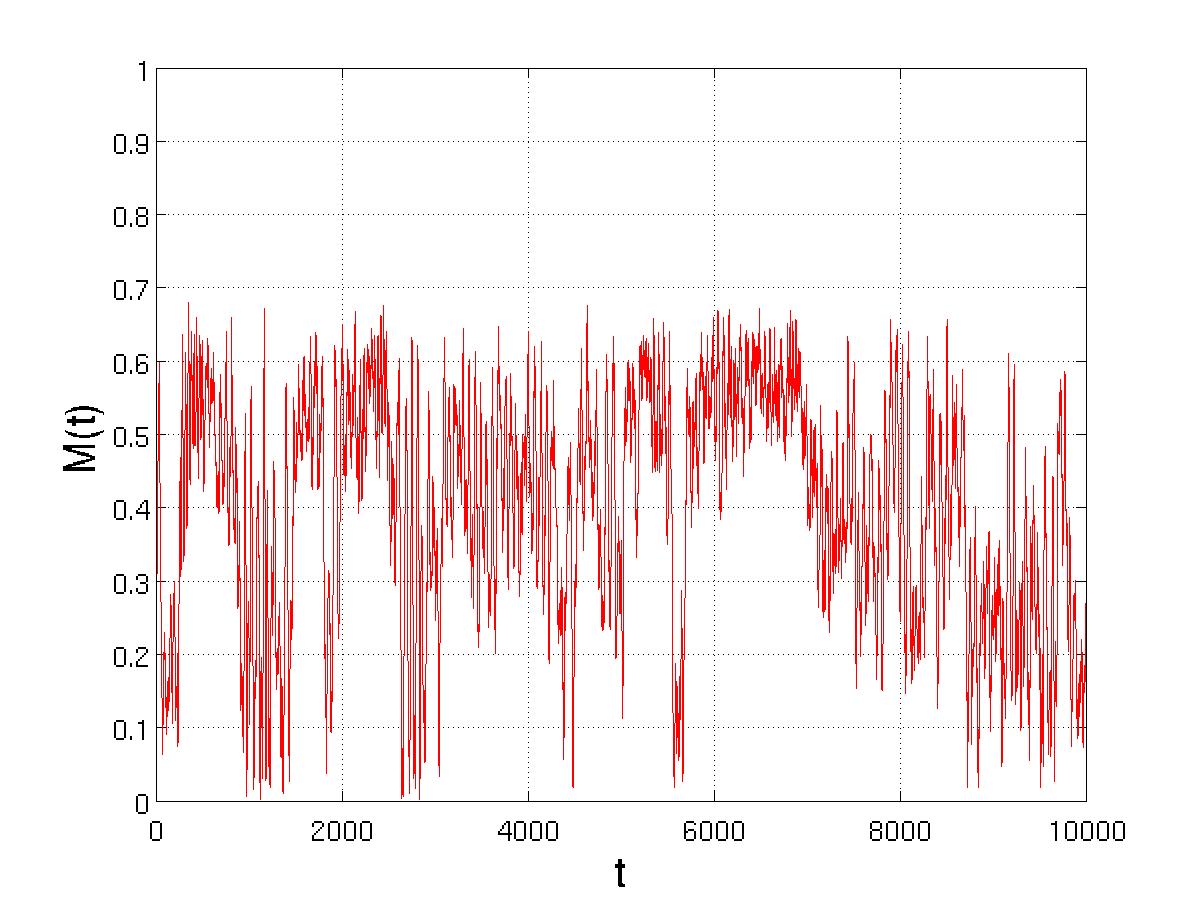}
\par\end{centering}

\centering{}\caption{Time evolution of the magnetization of the system with $\varepsilon\approx0.4$.
The system size is here $N=2^{9}$. The final time and data sampling
of the simulation is identical to the one performed in Fig.~\ref{fig:Fluctuations_Gamma}.
We notice that the fluctuations are indeed of the same order, however
we can notice that the typical time scale of the fluctuations appear
to be faster than in Fig.~\ref{fig:Fluctuations_Gamma}. \label{fig:Mag_evolution_512} }
 
\end{figure}
 In this section we study numerically the behavior of the order parameter
for different values of $\varepsilon$, $\gamma=1.5$ and different
system sizes with the aim of uncovering the timescales characterizing
the fluctuations. Indeed we can notice in Fig.~\ref{fig:Mag_evolution_512}
that the typical time scale of the fluctuations appears to depend
on the system size, as the magnetization fluctuations are much faster
for $N=512$ (Fig.~\ref{fig:Mag_evolution_512}) than for $N=8192$
(Fig.~\ref{fig:Fluctuations_Gamma}). Also, even though the signal
plotted in Fig.~\ref{fig:Mag_evolution_512} looks turbulent, it
may just be the consequence of the presence of a few unrelated modes.
\begin{figure}
\begin{centering}
\includegraphics[width=9cm]{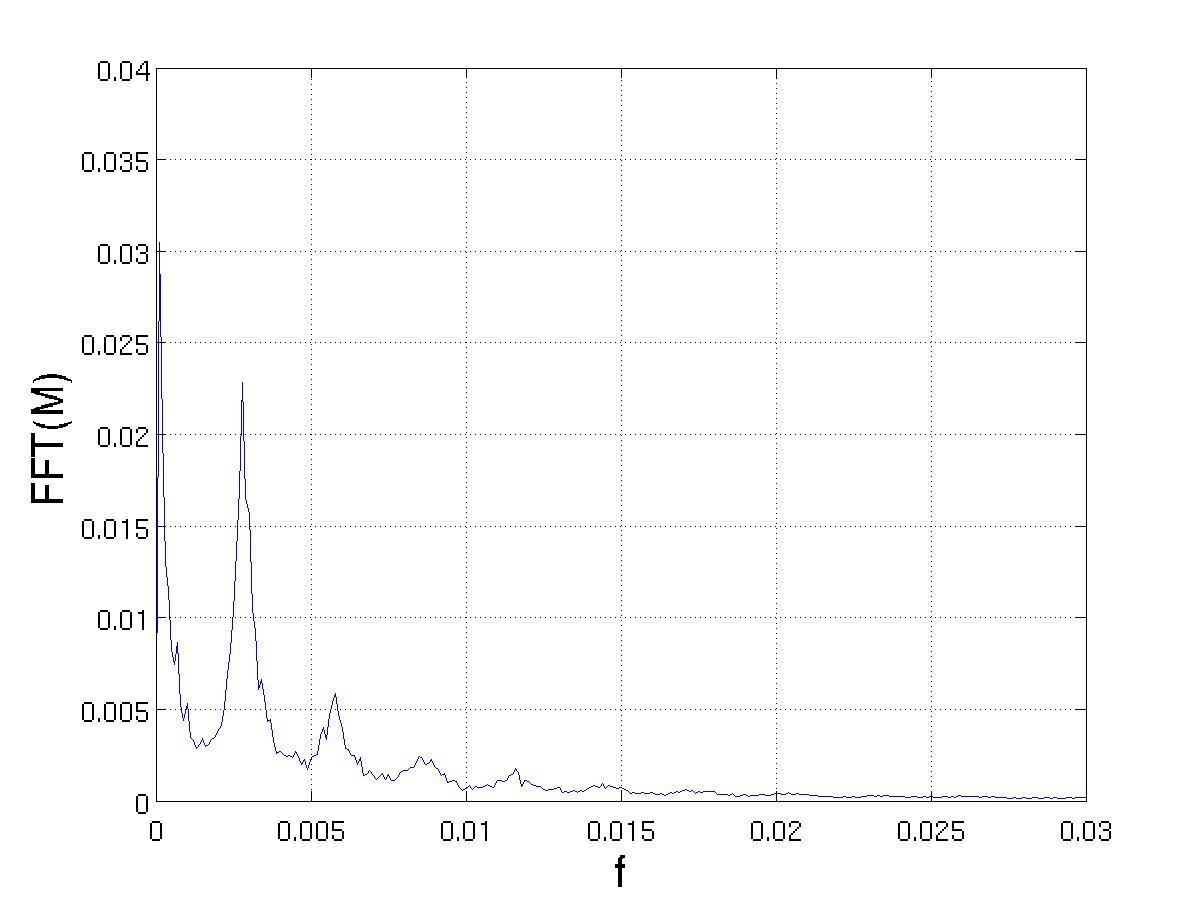}
\par\end{centering}

\centering{}\caption{Fourier spectrum of a ``chaotic'' signal of the order parameter.
The considered system size is $N=2^{14}$. We can notice that we obtain
a continuous spectrum with some broad peaks, and associated harmonics.
The time dependence is typically chaotic, and does not correspond
to a quasi-periodic signal.\label{fig:Fourier-spectrum}}
 
\end{figure}
In order to confirm the chaotic nature of the signal, we decided to
analyze its Fourier spectrum. An example of such spectrum is displayed
in Fig.~\ref{fig:Fourier-spectrum}. We can notice that the spectrum
is continuous, differently from the one given by a quasi-periodic
signal, so it is definitely of the chaotic (turbulent) type. However
we can notice as well some broad peaks which are associated to the
decreasing harmonics in this signal. Indeed, we can relate these peaks
to the typical scale of fluctuations that visually appeared in the
figures \ref{fig:Mag_evolution_512} and \ref{fig:Fluctuations_Gamma}.
\begin{figure}
\begin{centering}
\includegraphics[width=9cm]{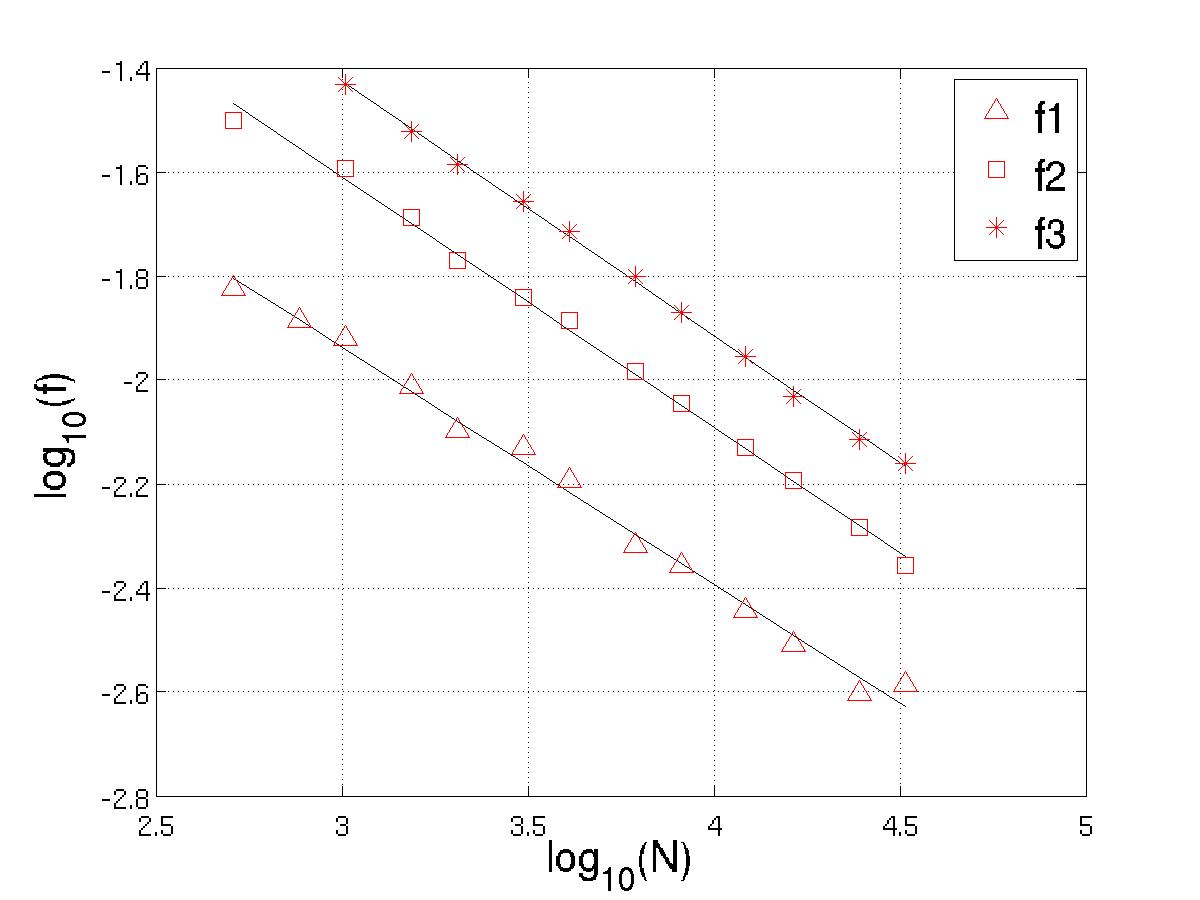}
\par\end{centering}

\centering{}\caption{Scaling of the localization of the frequency peaks versus system size.
The locations of the first three peaks (harmonics) are represented.
One notices a global uniform scaling of the different peaks, and a
slowing down of the typical fluctuation time. The scaling law shows
a decrease of the peak frequencies as $f\sim1/\sqrt{N}$.\label{fig:Scaling-of-the_frequency}}
 
\end{figure}
In order to determine the scaling with system size, we performed a
sequence of numerical simulations, with a fixed density of energy,
fixed total time and different system sizes. In these simulations,
the initial condition is extracted from a Gaussian distribution for
both the $p_{i}$'s and $q_{i}$'s. The signal analysis is performed
over the data that has been averaged during the second half of the
total simulation's time. The results are displayed in Fig.~\ref{fig:Scaling-of-the_frequency},
where the locations of the three first peaks displayed in Fig.~\ref{fig:Fourier-spectrum}
are represented versus system size in a log-log plot. One notices
a universal scaling of the typical fluctuation time scale, with all
peaks having a frequency that decreases as $f\sim N^{-1/2}$. This
scaling was initially not anticipated as one would naturally expect
a behavior similar to what has been observed for QSS's, with a typical
lifetime scaling $\tau\sim N^{\alpha}$, with $\alpha=1$ or higher
values. The observed scaling $\tau\sim N^{1/2}$ is another confirmation
that these chaotic states do not correspond to transient regimes,
but are ``steady''. We had already run very large time simulations
in without noticing any visible change in the dynamics of the order
parameter, but a transient with a large value of $\alpha$ could still
have been possible. 

We now move on to a theoretical hint at the observed scaling law,
and the confirmation as well that these are not transient states.

\subsection{Theoretical analysis}

In order to perform our analysis we carry out a similar calculation
as the one performed in \cite{DeNigris2013a}, that had allowed us
to prove that $\gamma=1.5$ was a threshold between the short range
and the long range behavior. The method was proposed in a general
context and explicitly developed for lattice's system in \cite{Leoncini2001}.
In order to be more self-consistent we review the method from the
start, and apply it to the considered system (\ref{eq: hamiltonian}). 

As already stated we consider a lattice (in dimension $D=1$ for our
system) of $N$ sites with coordinates $x_{i}=1,\cdots,N$. At each
site $i$ we have a particle, coupled to some neighbors, each having
a momentum $p_{i}$ and conjugate coordinate $q_{i}$. We recall that
we shall consider thermodynamical equilibrium properties (even though
we are looking at some dynamical properties) so the units are such
that the lattice spacing, the Boltzmann constant, and the mass are
equal to one. Also from the form of the Hamiltonian (\ref{eq: hamiltonian}),
a calculation within the canonical ensemble will imply that the $p_{i}$
are distributed according to a Gaussian distribution. Since we are
working on a lattice, with periodic boundary conditions, we can represent
the momentum as a superposition of Fourier modes: 
\begin{equation}
p_{i}=\sum_{k=0}^{Nk_{0}}\dot{\alpha}_{k}\cos(kx_{i}+\phi_{k})\:,\label{p}
\end{equation}
where the wavenumber $k$ is in the reciprocal lattice (an integer
multiple of $k_{0}=2\pi/N^{(1/D)}$), the wave amplitude is $\dot{\alpha}_{k}$,
and since we want the momenta to be Gaussian distributed variables
in the thermodynamic limit, we consider that the random phase $\phi_{k}$
is uniformly distributed on the circle. Therefore, we should, given
some conditions on the amplitude, be able to apply the central limit
theorem. The momentum set is labeled, using (\ref{p}), with the set
of phases $\ell\equiv\{\phi_{k}\}$. Note also that this equation
can also be interpreted as a change of variables, from $p$ to $\alpha$,
with constant Jacobian (the change is linear and we chose an equal
number of modes and particles).

Before proceeding, we would like to make some remarks. First, it is
clear from the Hamiltonian (\ref{eq: hamiltonian}) that we have a
translational invariance, which implies that the total momentum of
the system is conserved. Since physics should not change we make a
simple Galilean transform in order to choose a reference frame where
the total momentum is zero. The total momentum is directly linked
to the zero mode, so this choice implies thus that we have to take
$\dot{\alpha}_{0}=0$. Second, since we know that in the canonical
ensemble the variance of $p_{i}$ is fixed and equal to the temperature
of the system, we shall assume that the $\dot{\alpha}_{k}$ are all
of the same order (we need a large number of relevant modes for the
center-limit theorem to apply). Given these assumptions and using
the relation $\langle p_{i}^{2}\rangle=\sum\dot{\alpha}_{k}^{2}/2$
(we average over the random phases), we write $\langle p_{i}^{2}\rangle\approx T$
and obtain $\dot{\alpha}_{k}^{2}\approx O[(T/N)]$ (we call this relation
the Jeans condition \cite{Jeans_book}). 

We now move on to the associated conjugated variable of $p_{i}$,
since we have $\dot{q}_{i}=p_{i}$, we write it as 
\begin{equation}
q_{i}=\alpha_{0}+\sum_{k=k_{0}}^{Nk_{0}}\alpha_{k}\cos(kx_{i}+\phi_{k})\:,\label{q}
\end{equation}
where $\alpha_{0}$ is a constant since $\dot{\alpha}_{0}=0$, corresponding
to the constant average of the $q_{i}$'s. In order to proceed, since
we are below the mean-field critical temperature, we make a low temperature
hypothesis: thus, we can assume that neighboring $q_{i}$'s are not
too different although no global long range order exists. . Assuming
that the difference $q_{i}-q_{j}$ when the rotators interact is small,
we expand the Hamiltonian and obtain:
\[
H=\sum_{i}\frac{p_{i}^{2}}{2}+\frac{1}{4k}\sum_{i,j}\epsilon_{ij}(q_{i}-q_{j})^{2}.
\]
Using the previous expressions derived for $q_{i}$ and $p_{i}$ and
averaging over the random phases we end up with an effective Hamiltonian
\begin{equation}
\frac{\left\langle H\right\rangle }{N}=\frac{1}{2}\sum_{l=1}^{N}\dot{\alpha}_{l}^{2}+\alpha_{l}^{2}(1-\lambda_{l}),\label{alpha hamiltonian}
\end{equation}
where 
\begin{equation}
\lambda_{l}=\frac{2}{k}\sum_{m=1}^{k/2}\cos(\frac{2\pi ml}{N})\label{eq:eigen}
\end{equation}
are the eigenvalues of the adjacency matrix. 

We can extract from this a dispersion relation, indeed we have 
\begin{eqnarray}
\frac{d}{dt}(\frac{\partial\left\langle H\right\rangle }{\partial\dot{\alpha_{l}}}) & = & -\frac{\partial\left\langle H\right\rangle }{\partial\alpha_{l}}\nonumber \\
\ddot{\alpha_{l}} & = & -\omega_{l}^{2}\alpha_{l}\label{eq:Dispersion relation}
\end{eqnarray}
As mentioned this computation was already used in in order to show
that the critical threshold between short range and long range behavior
was $\gamma=1.5$; we used this formalism in order to compute analytically
the value of the magnetization in the thermodynamic limit. In the
present case, we stress the fact that the dispersion relation (\ref{eq:Dispersion relation})
embeds also some dynamical informations since we have access to the
typical frequencies that we can expect to find in the system. This
dynamical information was not used in previous papers levering this
formalism, but nevertheless the understanding of the observed scaling
law could provide new avenues for this approach. 

We can now use this dynamical feature in order to explain the critical
slowing down by monitoring how the spectrum behaves as we change the
size of the system, for the specific situation with $\gamma=1.5$,
i.e $k\sim\sqrt{N}$. For this purpose, we consider a specific mode
$l$; we have 
\begin{eqnarray}
\omega_{l}^{2} & = & (1-\lambda_{l})\label{eq:frequencies}\\
 & = & 1-\frac{1}{k}\left[\frac{\sin\left(\frac{\left(k+1\right)l\pi}{N}\right)}{\sin\left(\frac{l\pi}{N}\right)}-1\right]\:.
\end{eqnarray}
In order to proceed we shall consider that $N\rightarrow\infty$,
thus $N\gg\sqrt{N}$ , i.e $N\gg k$ and that $l$ is fixed, we can
then perform an expansion of the expression (\ref{eq:frequencies}),
and in order to avoid the first order $\omega_{l}^{2}=0$ result,
we shall expand it to third order using $\sin(x)=x-x^{3}/6+o(x^{3}).$
We then obtain (omitting the $o(x^{3})$ notation) 
\begin{eqnarray*}
\omega_{l}^{2} & \approx & \frac{k+1}{k}-\frac{1}{k}\frac{\frac{\left(k+1\right)l\pi}{N}-\frac{\left(k+1\right)^{3}l^{3}\pi^{3}}{6N^{3}}}{\frac{l\pi}{N}-\frac{l^{3}\pi^{3}}{6N^{3}}}\\
 & \approx & \frac{k+1}{k}\left[1-\frac{1-\frac{(k+1)^{2}l^{2}\pi^{2}}{6N^{2}}}{1-\frac{l^{2}\pi^{2}}{6N^{2}}}\right]\\
 & \approx & \frac{k+1}{k}\left[\frac{(k+1)^{2}l^{2}\pi^{2}}{6N^{2}}-\frac{l^{2}\pi^{2}}{6N^{2}}\right]\\
 & \sim & \frac{k^{2}}{N^{2}}\sim\frac{1}{N}\:.
\end{eqnarray*}
We recover analytically the critical slowing down exhibited numerically
in Fig.~\ref{fig:Scaling-of-the_frequency} and confirm that the
scaling law leads to $\omega\sim1/\sqrt{N}$, and thus characteristic
time scales of order $\sqrt{N}$.

\section{Concluding remarks}

In this paper we have analyzed the typical time scale $\tau(N$) over
which the chaotic behavior (fluctuations) of the order parameter evolves
as a function of system size. First after a numerical study, we have
exhibited that $\tau(N)\sim\sqrt{N}$. Then this behavior has been
afterwards confirmed theoretically, by showing that each of the frequencies,
associated to modes of the dual lattice, scaled as $\omega_{k}\sim1/\sqrt{N}$
with system size. The direct consequences of these results go in two
directions. First we confirmed the chaotic states observed and discussed
in \cite{DeNigris2013a,DeNigris2013b,DeNigris2015} indeed are not
a transient state like a QSS and, because of the presence of a large
continuous spectrum, we can as well confirm the chaotic nature of
the macroscopic behavior in these states, much like a turbulent one.
Second, when performing our theoretical analysis using the formalism
developed in \cite{Leoncini2001}, we were able to show for the first time that it is
possible to uncover some dynamical information from this formalism,
and the successful prediction of the scaling law shows that the formalism
is adequate to predict some finite size dynamical features of systems
with many degrees of freedom with underlying Hamiltonian microscopic
dynamics.

As a whole the typical decay of the characteristic time scale has
another important consequence:indeed should we consider an $N\rightarrow\infty$
limit, the fluctuations should stop and the system will end up frozen
in its initial magnetic state. It is important to comment that still
the infinite susceptibility would remain, so the system should remain
extremely sensitive to any external perturbation. This critical slowing
down with system size has been observed in other types of networks
with different structure. Thus, beside confirming the same behavior
arises considering lace networks as a substrate, an interesting perspective
would be to check if there are any similarities to what has been already
reported, and if this phenomenon could be of practical use, like for
instance to slow down the waves propagation in some localized regions.

\section*{acknowledgements}

S.D.N and X.L. would like to thank S. Ogawa  for fruitful discussions
and remarks during the preparation of this manuscript.

\end{document}